\documentclass[floatfix,a4paper,prl,twocolumn,superscriptaddress,preprintnumbers,amsmath,amssymb]{revtex4}

\usepackage{hyperref}
\usepackage[all]{hypcap} 
\usepackage{amsmath}
\usepackage{amssymb}
\usepackage{graphicx}
\usepackage{color}
\usepackage{multirow}
\usepackage{verbatim}

\begin{document}
\preprint{}

\title{Gate-controlled conductance enhancement from quantum Hall channels along graphene p-n junctions}

\author{Endre~T\'{o}v\'{a}ri}
\affiliation{Department of Physics, Budapest University of Technology and Economics, and Condensed Matter Research Group of the Hungarian Academy of Sciences, Budafoki \'{u}t 8, 1111 Budapest, Hungary}

\author{P\'{e}ter~Makk}
\affiliation{Department of Physics, University of Basel, Klingelbergstrasse 82, CH-4056 Basel, Switzerland}

\author{Ming-Hao~Liu}
\affiliation{Institut f\"{u}r Theoretische Physik, Universit\"{a}t Regensburg, D-93040 Regensburg, Germany}

\author{Peter~Rickhaus}
\affiliation{Department of Physics, University of Basel, Klingelbergstrasse 82, CH-4056 Basel, Switzerland}

\author{Zolt\'{a}n~Kov\'{a}cs-Krausz}
\affiliation{Faculty of Physics, Babes-Bolyai University, Str. Mihail Kogalniceanu nr. 1, 400084 Cluj-Napoca, Romania}
\affiliation{Department of Physics, Budapest University of Technology and Economics, and Condensed Matter Research Group of the Hungarian Academy of Sciences, Budafoki \'{u}t 8, 1111 Budapest, Hungary}

\author{Klaus~Richter}
\affiliation{Institut f\"{u}r Theoretische Physik, Universit\"{a}t Regensburg, D-93040 Regensburg, Germany}

\author{Christian~Sch\"{o}nenberger}
\affiliation{Department of Physics, University of Basel, Klingelbergstrasse 82, CH-4056 Basel, Switzerland}

\author{Szabolcs~Csonka}
\email{csonka@mono.eik.bme.hu}
\affiliation{Department of Physics, Budapest University of Technology and Economics, and Condensed Matter Research Group of the Hungarian Academy of Sciences, Budafoki \'{u}t 8, 1111 Budapest, Hungary}


\begin{abstract}
The formation of quantum Hall channels inside the bulk of graphene is studied using various contact and gate geometries. p-n junctions are created along the longitudinal direction of samples, and enhanced conductance is observed in the case of bipolar doping due to new conducting channels forming in the bulk, whose position, propagating direction and, in one geometry, coupling to electrodes are determined by the gate-controlled filling factor across the device. This effect could be exploited to probe the behavior and interaction of quantum Hall channels protected against uncontrolled scattering at the edges.
\end{abstract}

\maketitle

{\it Introduction---} The unique properties of graphene, such as the peculiar Berry phase leading to the half-integer quantum Hall effect\cite{Novoselov2005, Zhang2005}, the possibility to create p-n junctions, and the valley degree of freedom make it a versatile platform to study quantized conductance channels. Graphene can host spin and/or valley-polarized\cite{Zhang2006, Young2012, Yu2013, Abanin2013, Roy2014}, or fractional\cite{Bolotin2009, Du2009, Dean2011, Ki2014} quantum Hall channels, while appropriate engineering of the mechanical strain could lead to a quantum valley Hall effect\cite{Guinea2010, Low2010}. However, atomic scale disorder at the edges of a flake causes intervalley scattering, calling for an experimental platform where momentum-scattering is reduced, such as the nearly disorder-free environment of the bulk.

When a p-n junction is created across the width of a device, electron and hole-type quantum Hall channels (QHCs) that are usually located along the edges, copropagate along the junction. Cross-scattering between channels may equilibrate their current contributions, resulting in fractional values of the resistance quantum\cite{Williams2007, Abanin2007, Oezyilmaz2007, Ki2009, Lohmann2009, Velasco2012, Matsuo2015, Kumada2015}. In high-quality devices, current equilibration is diminished due to forbidden scattering between channels belonging to broken-symmetry Landau levels (LLs)\cite{Amet2014, Sanchez2016}, or to reduced disorder-broadening of levels and better spatial separation of channels\cite{Rickhaus2015}. For instance, the separation of QHCs is increased in Ref.~\citenum{Klimov2015} by using a softer potential step.

In the three different types of devices presented in this paper, quantum Hall channels are realized along the transport direction, between contacts, providing direct information on the conductance of the channels. The first is a two-terminal device with a p-n junction connecting source and drain, showing increased conductance when the filling factors of the two sides are opposite. The second one is of a similar design, but has two extra grounded terminals on the sides, allowing us to observe the current guiding effect of the p-n junction only. The third one has a bottom gate geometry that enables the formation of a circular p-n junction with tunable diameter and transmission to source and drain electrodes. Our results indicate that conducting channels are created in the bulk that are fully thermalized in the contacts like usual edge states, unaffected by the metal's doping and screening. We suggest using contact and local gate geometries that enable the formation of QHCs in the bulk - to preserve valley coherence - , and also the selective biasing of quantum Hall channels via appropriately placed grounding electrodes.

{\it Two-terminal p-n junction---} We have used a polymer-based suspension method following Refs.~\citenum{Tombros2011, Maurand2014} and a transfer method by Ref.~\citenum{Dean2010} for all three devices presented in this paper. Details are given at the end of the main text. Measurements were carried out at 1.5~K using low frequency lock-in technique. 

\begin{figure} [t]
\centering
\includegraphics[width=0.5\textwidth,clip]{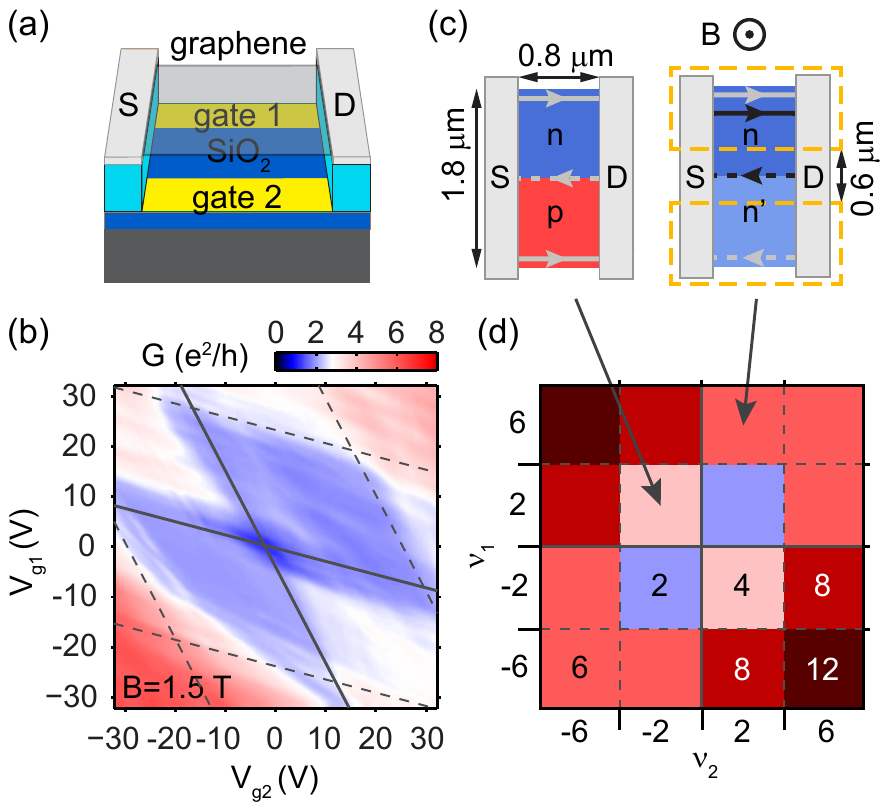}
\caption{(a) Structure of the first device: a graphene layer with two terminals, suspended over two bottom gates. (b) Conductance as a function of gate voltages at $B=1.5$~T, corrected for contact resistance. (c) Quantum Hall channel positions for bipolar (left) and unipolar (right) doping. Solid lines mark biased electron current trajectories, while dashed lines mark unbiased ones. Sample dimensions are indicated, with dashed orange lines showing the outlines of the bottom gates. (d) The expected conductance in units of $e^2/h$ as a function of filling factors $\nu_1,~\nu_2$. Color coding is the same as for (b).
}
\label{2term}
\end{figure}

A schematic of the first device is presented in Figure~\ref{2term}a. A single-layer graphene (SLG) flake is suspended between Pd source (S) and drain (D) electrodes, above two independently biased bottom gates. Figure~\ref{2term}b shows its differential conductance $G$ in units of the conductance quantum $e^2/h$ as a function of the gate voltages $V_\mathrm{g1},~V_\mathrm{g2}$, at a perpendicularly applied magnetic field of $B=1.5$~T. A checkerboard pattern emerges, where different regions - separated by dashed grey lines - mark the filling of different LLs. The distortion is the result of cross-capacitances\cite{Huard2007, Oezyilmaz2007, Gorbachev2008}. Solid grey lines distinguish the unipolar and bipolar quadrants.

Though the charge carrier density $n$ varies smoothly as a function of position, in order to visualize QHCs the average densities can be used to define the filling factors in the two halves of the flake: $\nu_{1,2}= n_{1,2}h / e B$. Along the diagonal of equipotential tuning ($V_\mathrm{g1}=V_\mathrm{g2}$), filling is uniform, and the expected quantum Hall plateaus\cite{Novoselov2005, Zhang2005} are observed near $2~e^2/h$. We extract an approximate serial contact resistance of $R_c \approx$1.4~k$\Omega$ from the plateau values. However, in the areas of bipolar doping, $G$ - corrected for $R_c$ - is increased to 3.5~$e^2/h$.

In the quantum Hall regime, Landau levels (LLs) form in the band structure of a two-dimensional conductor, exhibiting edge states of quantized conductance that propagate along the edges when an integer number of levels is filled in the bulk. In the case of non-uniform doping, the LL filling factor $\nu$ changes as a function of real-space position, and conducting channels may appear in the bulk of the sample. From here on, we refer to both types of propagating states as quantum Hall channels (QHCs). The conductance of our two-terminal device can be explained in the Landauer-B\"{u}ttiker formalism\cite{Buettiker1986, Buettiker1988, Beenakker1991} by picturing these channels. 

Fig.~\ref{2term}c shows two cases of non-uniform doping. The right panel is in the unipolar regime, with an n-n' junction near the center of the flake. In this example, the upper half has a filling of $\nu_1 = 6$ with two degenerate edge states at the top (grey lines from the 0th, and black ones from the 1st LL), while the bottom half of the flake has only $\nu_2 = 2$. Around the n-n' border, the filling changes, giving a fourfold degenerate QHC (black) of the 1st LL in the bulk. In an ideal sample, backscattering is absent, QHCs are fully thermalized at the contacts, and conductance is $G= \rm max (|\nu_1|,|\nu_2|) \cdot \it e^2/h$, determined by the number of biased channels (solid lines, from the source) counting all degeneracies. Dashed lines denote unbiased channels whose chemical potential is set by the drain to the global electrochemical potential. Fig.~\ref{2term}d depicts the expected plateau values in units of $e^2/h$ as a function of $\nu_1,~\nu_2$.

In the case of bipolar doping, as depicted in the left panel of Fig.~\ref{2term}c for the example of $\nu_2 = - \nu_1 = -2$, oppositely circulating states form in the two halves of the flake, with copropagating QHCs at the p-n interface. Ideally, conductance is given by the contribution of all channels connecting the source to the drain: $G=(|\nu_1| +| \nu_2|) \cdot e^2/h$, as displayed in Fig.~\ref{2term}d. After subtraction of $R_c$, the measured conductance (plotted in Fig.~\ref{2term}b) shows a maximum of $G \approx 3.5~e^2/h$ in the bipolar regime, which approaches the expected value of 4. It is most likely limited by backscattering between the channels in the bulk and at the edges, caused by residual disorder after current annealing of the sample. The enhanced conductance shows that new conducting channels are introduced in the bulk of graphene, despite the fact that contact electrodes partially screen the electrostatic potential of the gates, and also dope graphene, in their vicinity. However, we did not get direct information on where the current flows. To access the channels guided along the p-n interface, we have added further terminals to the design.

\begin{figure*} [t]
\centering
\includegraphics[width=1\textwidth,clip]{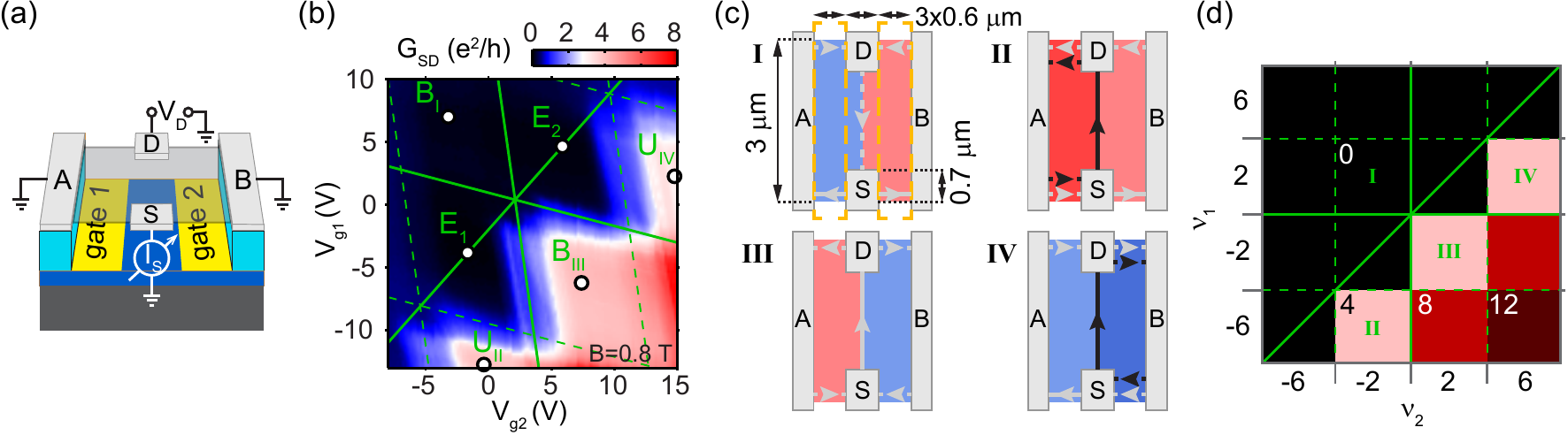}
\caption{(a) Setup of the second device, with current injected from and collected in the D and S electrodes, respectively, while contacts A and B are grounded. (b) Differential conductance between S and D as a function of the gate voltages at 0.8~T, corrected for a contact resistance of 1.2~k$\Omega$, which was estimated based on the expected plateau values shown in (d). Solid green lines separate areas of unipolar and bipolar doping, and mark the equipotential diagonal. Dashed lines distinguish areas of different filling factors. (c) QHCs in the electron injection picture from S, for various filling factor combinations $\nu_{1,2}$ of the left and right sides. Solid lines are biased electron channels, as opposed to dashed lines. Sample dimensions are indicated, with dashed orange lines showing the outlines of the bottom gates. (d) A map of the expected conductance as a function of $\nu_1,~\nu_2$, with Arabic numbers denoting the plateau values in units of $e^2/h$. Bold Roman numbers of different panels correspond to the examples of (c). Black-and-white circles in (b) mark points in unipolar (U) and bipolar (B) regions that correspond to the Roman-numbered cases in (c),(d), with a few points (E) along the equipotential diagonal.
}
\label{4term}
\end{figure*}

{\it Four-terminal p-n junction---} Fig.~\ref{4term}a shows the geometry of the second device. Here, an electrode (D) - situated above the gap between the gates - is biased by voltage $V_D$, and current $I_S$ is measured in a contact (S) on the opposite side. This is equivalent to the picture of injecting electrons from the source S with a chemical potential bias $eV_D$, and electron current measurement at drain D. Electrodes A and B on the left and right of the schematic ground all edge states, enabling us to study only the QHCs that propagate through the bulk. The conductance $G_\mathrm{SD}= \rm d \it I_S / \rm d \it V_D$ at 0.8~T, shown in Fig.~\ref{4term}b, exhibits the expected slanted checkerboard pattern as a function of the gate voltages. It drops below 0.04~$e^2/h$ at $\nu_1 = \nu_2 = \pm 2$, in the vicinity of points $E_1,~E_2$, while reaches a plateau of approximately $ 4 ~ e^2/h$ for $(\nu_1 ,\nu_2) = (-2,2)$ around point $B_\mathrm{III}$, as well as for $(\nu_1 ,\nu_2) = (-6,-2)$ ($U_\mathrm{II}$) and $(\nu_1 ,\nu_2) = (2,6)$ ($U_\mathrm{IV}$).

Most features can be explained in the Landauer-B\"{u}ttiker formalism. The (dashed) solid lines in Fig.~\ref{4term}c mark (un)biased electron channels, for various $\nu_1,~\nu_2$ filling factor combinations, while Fig.~\ref{4term}d shows the ideal plateau value of $G_\mathrm{SD}$ in units of the conductance quantum. Panels denoted by bold Roman numbers correspond to the cases in Fig.~\ref{4term}c. Depending on the sign and relation of $\nu_{1,2}$, we distinguish four regions on the map. (i) Along the equipotential diagonal $\nu_1 = \nu_2$, no direct channels exist between source and drain, and the injected electrons are fully absorbed in A and B. Above the diagonal, QHCs propagate from D to S, but since S is biased, $G_\mathrm{SD}=0$ (such as case \textbf{I}). (ii, iv) In the parts of the unipolar regions below the diagonal (like cases \textbf{II} and \textbf{IV}), a net electron current is carried from S to D by channels whose number is determined by the difference between the right and left filling factors: $G_\mathrm{SD} = | \nu_2 - \nu_1 | \cdot e^2/h$. (iii) In the bipolar quadrant below the diagonal (such as case \textbf{III}), all channels contribute to the current, and the conductance is $(|\nu_1| +|\nu_2|) \cdot e^2/h$. 

The measured plateaus of $4~e^2/h$ around point $B_\mathrm{III}$ in Fig.~\ref{4term}b matches the theory in Fig.~\ref{4term}d. Current flows directly from S to D, along the p-n junction, as depicted in Fig.~\ref{4term}c. Here we have measured the current flowing into contacts A and B as well, and found that approximately 89\% of the total electron current injected at S reaches D, suggesting that such p-n junctions can serve as high-efficiency electron guides. Widening source and drain contacts and the graphene flake, and increasing the magnetic field or the magnitude of the potential step across the junction may further increase the efficiency.

The plateaus of $G_\mathrm{SD} \approx 0$ at $\nu_1 = \nu_2 = \pm 2$ near points $E_1,~E_2$ of the equipotential diagonal are also in good agreement with expectations. Conductance at point $B_\mathrm{I}$ deviates slightly from the ideal value, possibly due to occasional scattering between the bottom and top edges of the flake, introducing finite electron current to D. We note that the plateau at $(\nu_1 ,\nu_2) = (2,6)$ (around $U_\mathrm{IV}$) is less developed than the one at $U_\mathrm{II}$, which can be attributed to a slight asymmetry in the annealed sample, resulting in nonzero transmission probability from the biased QHC (solid black line in panel \textbf{IV} of Fig.~\ref{4term}c) to the right-propagating (dashed black) channel at the top, bypassing the drain.

We have shown that in the vicinity of the Dirac-point, when LL occupation is $| \nu_1  = - \nu_2 | = 2$, a robust channel is formed in the bulk, acting as a direct, high-efficiency electron guide between source and drain. In the following, we investigate a more complex setup which allows us to study QHCs partially disconnected from the contacts in a circular geometry.

\begin{figure*} [t]
\centering
\includegraphics[width=1\textwidth,clip]{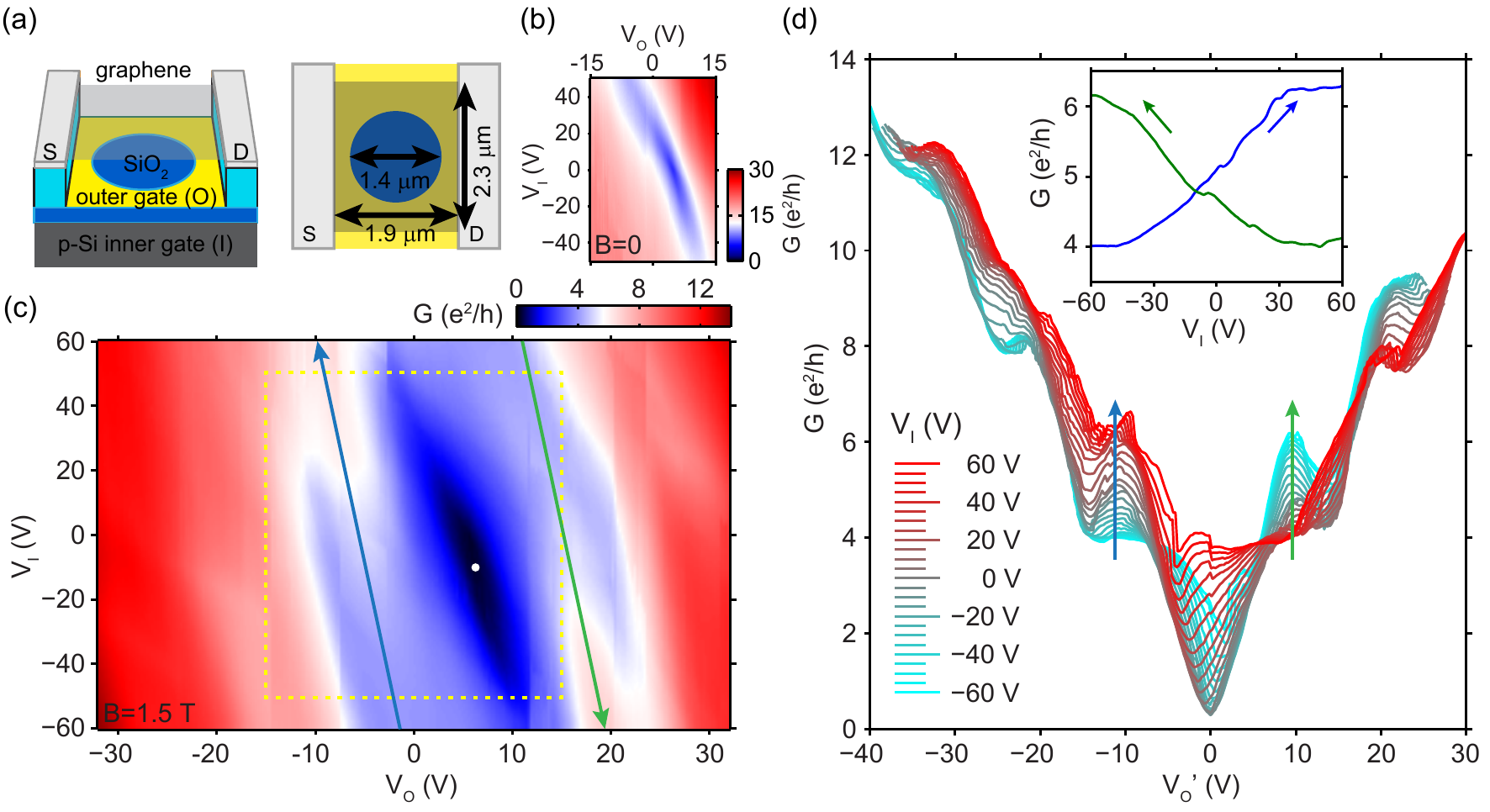}
\caption{(a) Schematic drawing with dimensions of the third, bilayer device tuned by a gate (yellow) with a hole, referred to as outer gate, and the Si backgate as inner gate. (b) Conductance $G$ as a function of the outer gate $V_O$ and inner gate $V_I$ voltages at $B=0$, and (c) at $B=$1.5~T. The dashed yellow rectangle in (c) highlights the gate voltage range used in (b). Both maps are corrected for $R_c \approx 0.42$~k$\Omega$ contact resistance. The white dot in (c) marks the point of minimum conductivity. (d) Conductance cuts at a series of $V_I$ voltages, horizontally shifted along $V_O$ by a linear function of $V_I$ to eliminate its cross-capacitance to the outer graphene areas. Inset: cuts along the two ridges of enhanced bipolar conductance, highlighted by blue and green arrows in (c) and (d). Hole side is corrected for $R_c\approx 0.42$~k$\Omega$, while electron side for $R_c=0$, in both (d) and its inset. 
}
\label{circ1}
\end{figure*}

{\it Circular p-n junction---} The third device we studied was a sheet of bilayer graphene, suspended over a bottom gate with a circular hole, as displayed in Fig.~\ref{circ1}a. The carrier density in the central part of the flake could be tuned through the hole by the doped Si backgate (referred to as the inner gate from here on) with bias $V_I$, while the surrounding area was doped by the bottom gate (later referred to as the outer gate) with voltage $V_O$. Two-terminal conductance $G (V_O,V_I)$ at zero $B$ field is depicted in Fig.~\ref{circ1}b. The data indicates that $V_I$ slowly moves the $V_O$ point of minimum conductance due to cross-capacitances, while increasing its value $G_\mathrm{min}$, for the Dirac-point is shifted inhomogeneously across the sample.

Fig.~\ref{circ1}c shows the conductance map at $B=1.5~$T. Quantum Hall plateaus of 4 and 8~$e^2/h$ appear in the unipolar regimes. A narrow region with a minimum conductivity of $0.3~e^2/h$ forms around the estimated Dirac-point (white dot), indicating that the 0th (zero-energy) LL starts to split into two fourfold degenerate levels due to electron-electron correlations\cite{Zhang2006, Young2012, Yu2013, Abanin2013, Roy2014}. 

The transition between the unipolar plateaus of 4 and 8~$e^2/h$ - see the lower left part of Fig.~\ref{circ1}c, with white color coding, parallel to the blue arrow - slowly moves as a function of $V_I$ due to the cross-capacitance between the inner gate and the outlying graphene regions. In order to eliminate this effect, we plot horizontal conductance cuts at a series of inner gate voltages in Fig.~\ref{circ1}d, all shifted along the $V_O$ axis by a linear function of $V_I$. As a result, the unipolar plateaus of the curves approximately overlap, and the blue and green arrows in Fig.~\ref{circ1}c correspond to those in Fig.~\ref{circ1}d. The electron side of the curves is corrected for $R_c=0$, while the hole side for $R_c=0.42~$k$\Omega$, to match expected plateau values of 4 and 8~$e^2/h$ at unipolar doping, found at $V_I \in [ -60, -40]$~V and $V_O' <0$, or $V_I \in [ 40, 60]$~V and $V_O' >0$.

The most striking features of the map in Fig.~\ref{circ1}c are ridges of enhanced conductance where one expects the bipolar regimes: at the upper part of the blue arrow, and the lower part of the green arrow. Figure~\ref{circ1}d shows that $G$ may be increased by more than $2~e^2/h$ with respect to the 4~$e^2/h$ plateaus, at $( V_O', V_I ) \approx (-11,60)~$V and $( V_O', V_I ) \approx (10,-60)~$V, respectively. The 8~$e^2/h$ plateaus are also enhanced in the bipolar regime. We suggest that the formation of new, circular channels in the bulk of graphene is the reason behind this conductance enhancement, whose contribution is not quantized due to partial transmission to contacts and to the overlap and scattering between the various QHCs. In the following, we discuss this concept in detail.

Since the outer gate screens a large part of the electrostatic potential of the Si inner gate, local normalized capacitance values $\rm d \it n (x,y) \rm / d \it V_{O,I}$ strongly depend on the real-space position $(x,y)$ on the flake. In order to get a qualitative picture of the formation and positions of quantum Hall channels, we have performed 3D electrostatic simulations on the electron density $n (x,y)$ for $B=0$.

\begin{figure} [t]
\centering
\includegraphics[width=0.5\textwidth,clip]{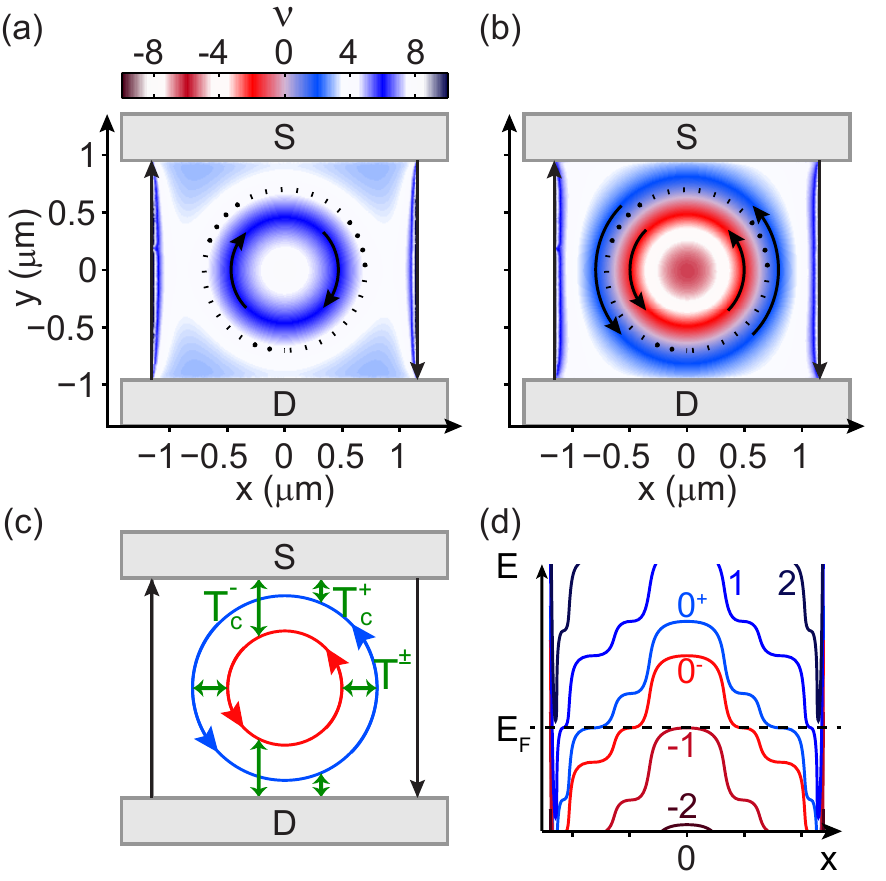}
\caption{(a) Zero-field electrostatic simulation of the electron density of the bilayer graphene flake, converted to filling factor using $\nu = n h / e B$, for $B=1.5~$T, at $\Delta V_O=9$~V, $\Delta V_I=50~$V from the Dirac-point, and (b) at $\Delta V_O=14$~V, $\Delta V_I=-50$~V. The dotted black line is the outline of the hole in the outer gate. Curved arrows mark the propagating directions of QHCs in the bulk, while straight arrows indicate usual edge states. (c) Structure of QHCs in the case of (b) and transmission possibilities between them and contacts. (d) Scheme of the Landau levels in (b) as a function of the $x$ coordinate, at $y=0$. LL numbering is defined by energy relations: $0^-$, $0^+$ originate from the originally zero-energy level, while 1 (-1) and 2 (-2) correspond to the first and second positive (negative) energy LLs.
}
\label{circ2}
\end{figure}

Fig.~\ref{circ2}a shows the Landau level filling factor $\nu (x,y) \propto n (x,y)$ across the bilayer flake for $B=1.5~$T, based on the simulated density map at $\Delta V_O = 9~$V and $\Delta V_I=50$~V from the Dirac-point, for unipolar electron doping. In the white regions of the map, an integer number of fourfold degenerate LLs is approximately full or empty, therefore they contain only localized states at the Fermi level. Although the first positive-energy LL of bilayer graphene is filled from empty to full in the highly doped central region (dark blue, $4< \nu < 8$), resulting in a circulating QHC whose propagation direction is given by black arrows, it does not contribute to current between the contacts or to backscattering between the edges, for they are insulated from each other by regions of integer filling (white). Despite charge accumulation near the edges\cite{Silvestrov2008, Cui2015}, conductance is determined only by usual edge states, indicated by straight arrows. As a result, this figure corresponds to a conductance of 4~$e^2/h$, qualitatively explaining the value of the green curve at large $V_I$ in the inset of Fig.~\ref{circ1}d.

Decreasing $V_I$ along the green line in Fig.~\ref{circ1}c keeps the density profile approximately constant in the outer parts of the flake, while it lowers the enhanced density in the center. The first LL is emptied, then, passing homogeneous doping, so is the 0th. Figure~\ref{circ2}b depicts the filling factor map at $\Delta V_O = 14~$V and $\Delta V_I=-50$~V from the Dirac point well into bipolar doping. The center of the flake is hole-doped: $\nu<-4$, indicating partial filling of the first negative-energy level. Following the $+x$ direction, an insulating region with $\nu \approx -4$ is crossed, then $\nu$ gradually increases to 4. 

As mentioned previously, the near-zero conductance in Figs.~\ref{circ1}c,d indicates that a gap is opening at the Dirac-point: the 0th LL splits into two fourfold degenerate levels, denoted by $0^-$ and $0^+$. Figure~\ref{circ2}d shows a sketch of the LL structure along a horizontal cross-section of the sample, consistent with the filling factor map in Fig.~\ref{circ2}b. The levels flatten when intersecting the Fermi energy $E_F$, for the density of states has a local maximum at the LL energy\cite{Martin2009, Yu2013}. 

Where the fourfold degenerate $0^-$ level is gradually filled with electrons (red stripe in Fig.~\ref{circ2}b), a circular propagating QHC forms, marked by an arrow. Further outside, the $0^+$ level is filled (blue stripe), again giving a QHC. The two states propagate in the same direction as in a regular p-n junction as a result of the slope of the LLs. Around $\nu = 0$, the Fermi-level is between the $0^-$ and $0^+$ levels, in a Landau gap. However, the fact that the sample exhibits a finite ($0.3~e^2/h$) conductance even when tuned homogeneously to this point (the white dot in Fig.~\ref{circ1}c) indicates that the disorder-broadened $0^-$ and $0^+$ levels still overlap, and the narrow region of $\nu \approx 0$ between the QHCs of the levels is not insulating. 

Fig.~\ref{circ2}b suggests that the channel belonging to the $0^+$ level (blue) has finite transmission to the contacts. Consequently, the inner and outer circular QHCs on the sides of the p-n junction act as extra current-carrying states between source and drain, and give a positive contribution to the base conductance of 4~$e^2/h$ of the edge states. Thus the simulation in Fig.~\ref{circ2}b qualitatively corresponds to the enhanced-conductance ($V_I = -60~$V) end of the green line in the inset of Fig.~\ref{circ1}d.

Based on the electrostatics in Fig.~\ref{circ2}b, Fig.~\ref{circ2}c shows the structure of the circular propagating channels with possible transmissions between them and the contacts. $T_c$ transmission probabilities indicate scattering mechanisms from the QHCs of the $0^-,~0^+$ levels to the contacts, and $T^{\pm}$ to each other. Backscattering between the circular channels and the edge states is most likely negligible, for they are insulated by a region of near-integer filling. 

In a simple example with realistic assumptions, we estimate the conductance contribution of the circular QHCs. Ideally, the outer channel is fully transmitted to the contacts, while the inner one is most likely too far away, and fully reflected. Thus, $T^+_c=1$, and $T^-_c=0$. Since the inner and outer channels overlap (for the $0^- - 0^+$ Landau gap is not well-developed), current that is injected only to the outer channel at the source is distributed between them. At the opposite contact, only the outer channel's current is drained. Considering that both channels are fourfold degenerate, and assuming full equilibration along their trajectory, their conductance enhancement is $\Delta G = 2. \dot{6}~e^2/h$. This value is slightly larger than the observed $\sim 2.2~e^2/h$. The enhancement may be limited by backscattering to edge states or imperfect coupling to electrodes. In contrast, if the gap between the $0^-$ and $0^+$ levels was well formed, the enhancement would be higher, up to the maximum possible contribution of the outer channel, $4~e^2/h$.

If we slightly raise the voltage of the outer gate, the density increases in the graphene areas above it. The circular blue stripe ($\nu \sim 2$) in Fig.~\ref{circ2}b shrinks, the QHC of the $0^+$ level becomes insulated from the contacts, and a local conductance minimum is expected to appear, in agreement with measurements in Figs.~\ref{circ1}c,d.

Along the green line of Fig.~\ref{circ1}c, the filling factor profile of the outer parts of the flake remains approximately constant. Decreasing $V_I$ continuously changes the doping of the central part from electron to hole. In a range of $V_I$ values, the $0^+$ LL is partially filled with electrons in most parts of the flake and conducts diffusively between the source and drain electrodes. Due to the larger-than-one aspect ratio of the device, this may be the reason for increased conductance\cite{2term_Abanin2008, 2term_Williams2009} that is observable already in the unipolar regime (inset of Fig.~\ref{circ1}d). Further decreasing $V_I$, this local conductance maximum evolves into the ridge of enhanced conductance in Fig.~\ref{circ1}c. This monotonous transition can be explained by the formation of the $0^+$ level's circular QHC, and the gradual increase in its diameter, resulting in better and better $T^+_c$ coupling to the contacts. The formation of a plateau around $6.2~e^2/h$ in the measured conductance suggests that $T^+_c$ eventually reaches close to unity transmission. The evolution of the blue line of Fig.~\ref{circ1}c is caused by the same mechanism, but with opposite signs of the filling factors.

The same effect can be seen at higher plateaus: the electron (hole) side unipolar $8~e^2/h$ plateau's conductance also increases in the bipolar regime. In this case, it is the 1st (-1st) LL that forms a circular QHC coupled to the contacts, enhancing the conductance. However, channels are more tightly packed and the insulating regions are narrower, for the density gradient is higher. Scattering between circular and edge states is increased, consequently, their contribution is somewhat smaller than for lower plateaus.

Besides the device shown in Fig.~\ref{circ1}a, we have performed control measurements on another, single-layer sample with a holey outer gate, where the hole diameter was 1~$\mu$m, the width 1.4~$\mu$m, while the source-drain distance remained almost the same, 1.8~$\mu$m. Here, the contacts were located 400~nm, and the flake edges 200~nm from the hole's border in plan view, compared to the 250~nm and 450~nm values, respectively, of the bilayer flake described above. No positive or negative change was observed in the $2,~6~e^2/h$ plateaus in the same voltage range, suggesting that the increased screening of the outer gate decreased the size of the inner gate induced circular QHCs enough that they were fully decoupled from the contacts, as well as from the edge states.

{\it Conclusions---} We have examined three types of local gated samples. Measurements on the two and four-terminal devices prove that Hall channels propagating along a p-n junction can be fully absorbed in a contact despite its screening and doping, and contribute to conductance in a quantized way. Our results show that p-n junctions can serve as high-efficiency current guides, and indicate that different Landau levels' co-propagating edge states can be detached from the edges by local gating and independently biased using grounded contact electrodes, suggesting a way to study the physics of spin and valley-polarized, or fractional channels avoiding disorder and valley decoherence at edges. This is a huge advantage, since although interesting phenomena like the formation of valley-polarized edge states are predicted using strain\cite{Guinea2010, Low2010} in properly engineered suspended graphene, the atomically rough edges would inevitably cause scattering between these channels. 

Moreover, circularly propagating quantum Hall channels have been created, whose size and coupling to contacts depend on the gate voltages. These observations demonstrate the ability to tune a propagating channel's trajectory such that transmission to electrodes or other channels is controlled, paving the way for graphene quantum point contacts and interferometers operated in the quantum Hall regime: experiments that, so far, have been available only in 2D semiconductor systems\cite{Chamon1997, Mueller1992, De-Picciotto1997}.

\bigskip

{\it Acknowledgements---} We acknowledge useful discussions with Romain Maurand, Andreas Baumgartner, Andr\'{a}s P\'{a}lyi, P\'{e}ter Rakyta, L\'{a}szl\'{o} Oroszl\'{a}ny, Matthias Droth, and Csaba T\H{o}ke. This work was funded by the EU ERC CooPairEnt 258789, Hungarian Grants No. OTKA K112918, and also the Swiss National Science Foundation, the Swiss Nanoscience Institute, the Swiss NCCR QSIT, the ERC Advanced Investigator Grant QUEST, the Flag ERA iSpinText, and the EU Graphene Flagship project. M.-H. L. and K. R. acknowledge funding from Deutsche Forschungsgemeinschaft within SFB 689.

\bigskip

{\it Methods---} Fabrication steps followed Refs.~\citenum{Tombros2011, Maurand2014}. First, 5/45-55~nm thick Ti/Au bottom gates were fabricated on a p:Si wafer covered by 300~nm SiO$_{2}$, which were covered first with an electron-beam evaporated, 40~nm thick MgO insulating layer (not displayed in the figures), second with 600~nm thick LOR resist. Graphene was exfoliated onto a separate wafer and transferred using the method described in Ref.~\citenum{Dean2010}. Subsequently, the flake was contacted with 40 or 60~nm thick Pd wires, and etched using e-beam lithography and reactive ion etching. Finally, graphene was suspended by exposing and developing the LOR resist below. To remove solvent and polymer residues, samples were current annealed at 1.5~K in a vacuum. Measurements were performed in the same conditions, using standard lock-in technique.

The 3D electrostatic model is built on the device dimensions in Fig.~\ref{circ1}a, and is used to obtain the self-partial capacitances \cite{Cheng_EM_1989, Liu2013} to individual metal contacts and gates, via the finite-element simulator FENICS\cite{Logg2012} combined with the mesh generator GMSH\cite{Geuzaine2009}. Electron density maps were calculated at zero magnetic field, and in a not self-consistent way, i.e. without taking into account the formation and screening effects of the compressible areas, where a Landau level is partially filled. In spite of this limitation, we have achieved a good qualitative representation of the circular QHCs.






\end{document}